# Thermal and Dielectric Properties of Juno's Regolith at One Millimeter Wavelength

Jian-Yang Li (李荐扬)[1*], Timothy N. Titus[2], Arielle Moullet[3], Henry H. Hsieh[1]

[1] Planetary Science Institute, 1700 E. Ft. Lowell Rd., Tucson, AZ 85719, USA
[2] US Geological Survey, Flagstaff, AZ, USA
[3] SOFIA/USRA, Moffett Field, CA, USA

## Abstract

We present the modeling results of the thermal lightcurve of asteroid (3) Juno at the wavelength of $\lambda = 1.3$ mm measured by the Atacama Large Millimeter-submillimeter Array. A thermophysical model together with a radiative transfer model suggest a thermal inertia of 13±10 [J m$^{-2}$ K$^{-1}$ s$^{-0.5}$], an equivalent emissivity of 0.8±0.1, a loss tangent of 0.4±0.3, and an index of refraction 1.8±0.3. Based on previous laboratory measurements, the modeled index of refraction suggests a regolith porosity of about 45%. However, thermal inertia models using the material parameters of ordinary chondrite indicate a grain size of 10s µm and require a high porosity of ~90% to explain the low thermal inertia. In order to explain such a contradiction, we postulate that some repulsive mechanism might be in effect to reduce the contact of grains and therefore the thermal inertia. The loss tangent of Juno's regolith corrected for the modeled thermal skin depth is in the order of 0.5, much higher than that of the lunar regolith and indicating an electrical skin depth of $L = 0.1 – 1.4$ mm that is within the thermal skin depth. The shape of the rotational lightcurve of Juno in the mm wavelengths is dominated by its irregular shape, but rotational variations in the thermal and/or dielectric properties cannot be ruled out. Our results demonstrate that mm-wavelength observations of asteroids provide an extra dimension of constraints to the porosity and grain size of asteroid regolith compared to the thermal infrared observations.

## 1. Introduction

Thermal observations of asteroids at mm to sub-mm wavelengths not only reveal the temperature characteristics of the surface, but also probe subsurface regolith properties. For particulate regolith on asteroids, porosity usually dominates the thermal conductivity over composition (Keihm et al. 1973, Okada et al. 2020, Cambioni et al. 2021). In addition, the depth probed by such observations depends on the bulk dielectric properties of the subsurface material, which is also strongly affected by porosity (e.g., Brouet et al. 2015). Therefore, mm-wavelength observations of asteroids have the potential to provide unique constraints about the physical properties of asteroidal surfaces and subsurfaces, such as thermal inertia, porosity. Historically, millimeter and submillimeter observations have been performed for large asteroids but in a disk-integrated sense due to the limited spatial resolution available for the observing facilities (e.g., Redman et al. 1998, Gulkis et al. 2010, 2012). The Atacama Large Millimeter-submillimeter Array (ALMA) enabled mapping of the surface thermal and dielectric properties of large asteroids with implications for their surface geological processes (e.g., de Kleer et al. 2021, 2024, Cambioni et al. 2022, Simpson et al. 2022, Phua et al. 2024). The thermal properties of asteroid regolith are directly relevant to the regolith formation and loss process (e.g., Harris and Drube 2016, MacLennan and Emery 2022).

---

[*] Current address: Planetary Environmental and Astrobiological Research Laboratory, School of Atmospheric Sciences, Sun Yat-sen University, Zhuhai, China. lijianyang@mail.sysu.edu.cn

Here we present the modeling of the flux of asteroid (3) Juno at a wavelength of λ = 1.3 mm using data obtained by ALMA as part of its first long-baseline campaign (ALMA Partnership et al., 2015). Juno has a diameter of ~250 km with a shape described as "lozenge-like", indicating that it may lie in the size regime between largely spherical dwarf planets and more irregularly shaped large- to medium-sized asteroids (Viikinkoski et al. 2015). The asteroid was observed by ALMA for 4.4 hours in October 2014, covering 60% of its rotational period $P$ = 7.2 hours and resulting in ten images (ALMA Partnership et al. 2015). The thermal flux of Juno varies with its rotational phase, dominated by its non-spherical shape (Viikinkoski et al. 2015). We focused on the disk-integrated thermal flux and modeled the diurnal temperature of Juno with a thermophysical model coupled with a radiative transfer model to account for the subsurface thermal emission, following Li et al. (2020). Thermal and dielectric properties were constrained by the modeling fitting, allowing us to infer the physical properties of Juno's regolith in the top few mm depths.

## 2. Data and Modeling Results

ALMA Partnership et al. (2015) reported that ALMA observations of Juno's thermal continuum emission were obtained on 2014 October 19 at a central wavelength of λ = 1.29 mm (frequency $f$ = 233 GHz) for a total of ~5 hours. After the standard data reduction and flux calibration, they reconstructed ten images, each covering a duration of 10 – 18 min. The total thermal flux density, $F$, was measured and reported from all images. We based our thermal modeling analysis on those mm-wavelength flux densities.

Our thermal and radiative transfer modeling follows Li et al. (2020). In thermophysical models (e.g., Spencer et al. 1989), surface and subsurface temperatures of a planetary body are determined by thermal inertia, which is defined as $\Gamma = \sqrt{\kappa \rho c}$, where κ is thermal conductivity, ρ is bulk density, and $c$ is the specific heat capacity. Thermal inertia measures how effectively a material resists changes in temperature induced by heat transfer. The surface of a planetary body undergoes diurnal and seasonal variations of temperature in response to periodic solar insolation. The amplitude of temperature variations diminishes with increasing depth beneath the surface, and the delay of peak temperature relative to the maximum local insolation increases, both depending on thermal inertia. The depth at which the amplitude of a periodic thermal wave is diminished by a factor of 1/$e$ is defined as thermal skin depth,

$$\delta = \frac{\Gamma}{\rho c} \sqrt{\frac{P}{\pi}} \qquad (1)$$

with *P* being the period of thermal wave. Generally, at depths below several thermal skin depths, the temperature is considered stable and no longer responsive to temporal changes in insolation.

Empirical thermal modeling in the IR wavelengths (e.g., Lebofsky et al. 1986, Harris 1998) usually includes a "beaming" parameter to describe how much the observed thermal radiation from an object deviates from that of a spherical object of the same effective size with a smooth surface under an instantaneous thermal equilibrium (i.e., $\Gamma = 0$). This deviation is interpreted as being due to an object's non-zero thermal inertia that changes the surface temperature distribution from the nominal $\Gamma = 0$ case (Delbo et al. 2007), and roughness that changes the daily insolation profile and causes self-heating (Emery et al. 1998). In thermophysical modeling, the empirical beaming parameter is not needed. Instead, we use an equivalent emissivity, $\varepsilon_{eq}$, in the thermal balance equation and let it vary to account for the effect of roughness (Capria et al. 2014). Compared to the more complicated thermophysical model that includes craters or statistical roughness model for surface roughness (e.g., Rozitis & Green 2011, Warren et al. 2019), the approach that we adopt here is much more computationally feasible, given that a radiative transfer model with additional parameters must be incorporated (see later text) to model the sub-mm data.

In our calculations, we assumed uniform albedo and thermal properties across Juno's surface and used a single-layer model with no vertical gradient in thermal and dielectric properties. We adopted the triangular plate shape model and the pole orientation of Juno (Viikinkoski et al. 2015) as archived in the Database of Asteroid Models from Inversion Techniques (DAMIT[1]; Ďurech et al. 2010) to represent the irregular shape and rotation of Juno. Diurnal surface and subsurface temperatures were calculated with Arizona State University KRC code (Kieffer 2013; where the KRC acronym refers to the terms for conductivity, density, and specific heat in the definition of thermal inertia) based on the thermophysical model at the orbital geometry corresponding to the ALMA observations (heliocentric distance of $r_h = 2.07$ au and a geocentric distance of $\Delta = 1.97$ au). The bolometric Bond albedo of Juno was assumed to be $A = 0.09$ (ALMA Partnership et al. 2015). The other parameters for KRC calculation were assumed to be $\rho = 1300$ kg m$^{-3}$, and $c = 630$ J K$^{-1}$ kg$^{-1}$. These assumed values provide the initial scaling for the thermal skin depth, which will be adjusted based on the asteroid models to derive the density of the regolith (see Section 3.1).

Once the temperature distribution on Juno's shape model and in depth was established, we used a radiative transfer model (Li et al. 2020) to calculate the modeled flux density from the subsurface. The mm-wavelength observations effectively probe the temperature beneath the surface characterized by the electric skin depths, *L*, which represents the radiative transfer optical depth inside a medium. The electric skin depth is the reciprocal of the electric absorption coefficient,

$$L^{-1} = \mu = \frac{4\pi n}{\lambda} \sqrt{\frac{\sqrt{1+\tan^2 \Delta} - 1}{2}} \qquad (2)$$

where $\tan \Delta = K_{ei}/K_{er}$ is the loss tangent of the medium, with $K_{er}$ and $K_{ei}$ being the real and imaginary parts of the complex dielectric constant, $K_e$, respectively. The index of refraction, *n*, is related to the real part of the complex dielectric constant, $n \approx \sqrt{K_{er}}$. Note that in this equation, we assumed $k/n \ll 1$, where *k* is the imaginary part of the complex index of refraction. Given that $\tan \Delta = \frac{2nk}{n^2 - k^2} = 2\frac{k}{n} / \left(1 - \frac{k}{n}\right)^2$, this assumption implies $\tan \Delta \ll 1$. We will return to this assumption when we discuss the model fitting results in Section 3.1.3.

---
[1] https://astro.troja.mff.cuni.cz/projects/damit/

With the radiative transfer model, the thermal emission is integrated through the subsurface along the optical path. Assuming electric skin depth is independent of temperature (and therefore depth), the integrated emission $I_\nu$ is,

$$I_\nu = (1 - R_\nu) \int_0^\infty I_\nu(T(z)) \exp\left(-\frac{z}{L \cos i}\right) \frac{dz}{L \cos i} \qquad (3)$$

where $T$ is temperature at depth $z$; $i$ is the incidence angle of thermal emission before it crosses the surface boundary, which is related to the emission angle $e$ by Snell's refraction law and depends on the index of refraction, $n$; $R_\nu$ is the Fresnel refraction coefficient, which depends on $i$ and $n$ and is the average of the coefficients of the two perpendicular polarizations. The total model flux density was integrated from the simulated images of Juno.

Fig. 1 shows simulated lightcurves of Juno using various $\Gamma$ and $\tan \Delta$ values. These models clearly show that the rotational variation of Juno's $\lambda$ = 1.3 mm flux density is dominated by its non-spherical shape. The simulated lightcurve amplitude is about $\Delta F$ = 20 mJy, or 10% of the average flux, $F_{avg}$. All modeled lightcurves are double peaked with the two maxima differing by 7 mJy (3.6% $F_{avg}$) and two minima differing by 13 mJy (13% $F_{avg}$). Different thermal inertias result in different maxima and minima values varying by 2 – 3 mJy, which is much smaller than the lightcurve amplitude or the difference between the two maxima or the two minima. On the other hand, different thermal inertias will cause the longitudes of lightcurve extrema to shift by up to ~30º, depending on the loss tangent. For example, when $\tan\Delta = 0.32$, the lightcurve minimum is at ~150º longitude when $\Gamma < 40$ [J m$^{-2}$ K$^{-1}$ s$^{-0.5}$] (hereafter "thermal inertia unit", or tiu), but shifts to ~120º when $\Gamma = 640$ tiu. Therefore, the thermal properties of Juno are best constrained by modeling the overall shape of the observed thermal lightcurve rather than observations at a single snapshot at a particular longitude, even when a shape model is available and the rotational phase can be precisely determined.

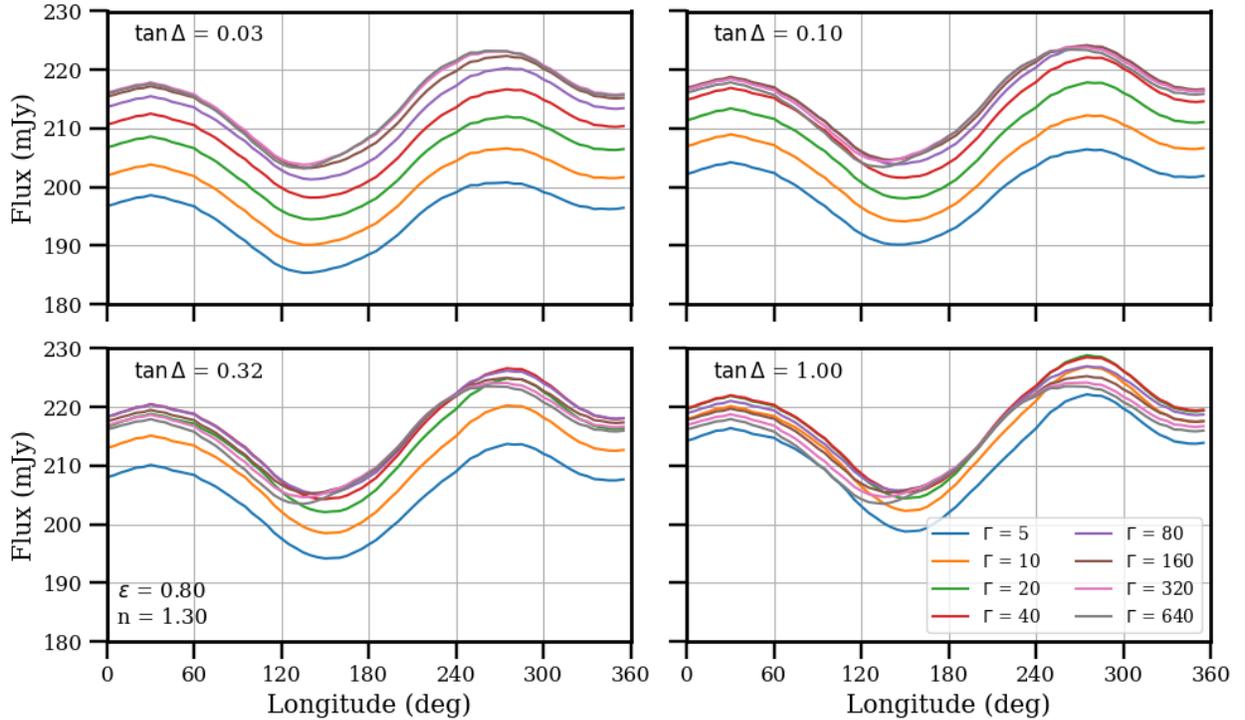

**Figure 1.** Model lightcurves of Juno with various thermal inertias as listed in the legend and different loss tangents for each panel. The emissivity and index of refraction are fixed to the values marked in the lower left panel. The overall shape of lightcurves is dominated by the shape of Juno, whereas the absolute temperature scale, the amplitude, and the longitudes of extrema of the lightcurves depend on the thermal and dielectric parameters.

To retrieve the best-fit parameters, we performed model calculations covering a grid of parameter space, including eight thermal inertia values from $\Gamma = 5$ to 640 tiu in logarithmic intervals, four equivalent emissivity values from $\varepsilon_{eq} = 0.7$ to 1.0, nine loss tangent values from $\tan\Delta = 1\times10^{-4}$ to 1.0 in logarithmic intervals, and eleven index of refraction values from $n = 1.001$ to 2.5 in uniform linear intervals. The ranges of the parameters covered in our modeling were based on the typical ranges of values derived from thermal analysis of asteroids of similar size and taxonomical type as Juno (e.g., Delbo et al. 2015, MacLennan & Emery 2022, Masiero et al. 2011). Output modeled flux densities, $F_{model}$, were compared with all ten flux density measurements of Juno, $F_{measure}$, and $\chi^2$ values were calculated as,

$$\chi^2 = \sum_i \left(\frac{F_{i,measure} - F_{i,model}}{\sigma_i}\right)^2 \qquad (4)$$

where $\sigma_i$ is the reported flux density uncertainties (ALMA Partnership et al. 2015). The reduced $\chi^2$ is defined as $\chi^2_{red} = \frac{\chi^2}{N-4}$, where $N = 10$ is the number of data points, and 4 is the number of parameters in the fit. To assess the goodness of fit and the uncertainty of model parameters, we adopted all 19 solutions that satisfy $\chi^2_{red} < \chi^2_{red,min}\left(1 + \sqrt{\frac{2}{N-4}}\right)$ (e.g., Press et al. 1986), where $\chi^2_{red,min}$ is the $\chi^2_{red}$ of the best-fit model, as indistinguishable solutions (Figs. 2, 3, Table A1) for further analysis.

Overall, the best-fit model lightcurve matches the data reasonably well, with a root-mean-square residual of 1.9 mJy, or 1.0% $F_{avg}$. The largest deviations are near the lightcurve minimum around a longitude of 180º, where observed flux density is lower than the model by about 2%. The average parameters of the top 19 acceptable models, weighted by $1/\chi^2_{red}$, are $\Gamma = 13\pm10$ tiu, $\varepsilon_{eq} = 0.8\pm0.1$, $n = 1.8\pm0.3$, $\tan\Delta = 0.4\pm0.3$. The uncertainties are simply the weighted standard deviation of the fitted parameters, not necessarily representing the statistical confidence level of the best-fit values. The distributions of the top 19 best-fit solution parameters are shown in Fig. 3. The distribution of thermal inertia has a clear upper boundary, whereas the lower boundary is not well defined. Given the lower end of 5 tiu for the values that we modeled, it is safe to say that the thermal inertia of Juno is <25 tiu, and the likelihood for a thermal inertia of higher than the upper bound of 640 tiu for our model is low. The equivalent emissivity shows a quite flat distribution, indicating that we do not have a strong constraint on this parameter, although a lower bound between 0.6 and 0.7 seems likely. The distribution of the index of refraction shows a peak near 1.7 with a sharp lower bound and a relatively shallow upper bound. The range seems to be consistent with the mean and standard deviation. The loss tangent shows a clear peak near 0.3 with a range of 0.1 to slightly less than 1, again consistent with the mean and the standard deviation given above.

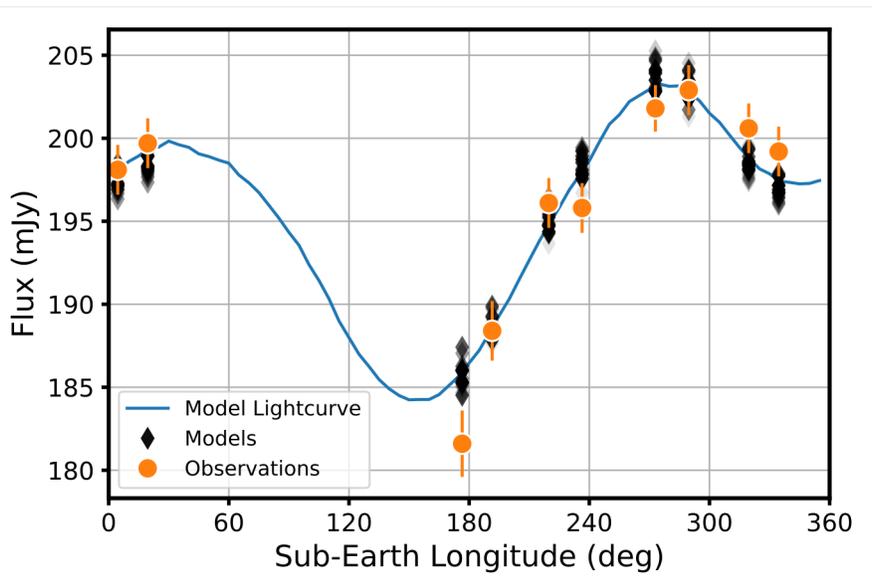

**Figure 2.** Best-fit model to the observed λ=1.3 mm flux of Juno. The orange circles with error bars are the observed flux densities (ALMA Partnership et al. 2015). The black diamonds of various shades are the top 19 best-fit models that are statistically indistinguishable, with darker symbols representing better fit. The solid curve is the lightcurve corresponding to the best-fit model. The best-fit $\chi^2_{red}$ = 2.0, and the corresponding RMS is 1.9 mJy, ~1.0% of the average observed fluxes, $F_{avg}$.

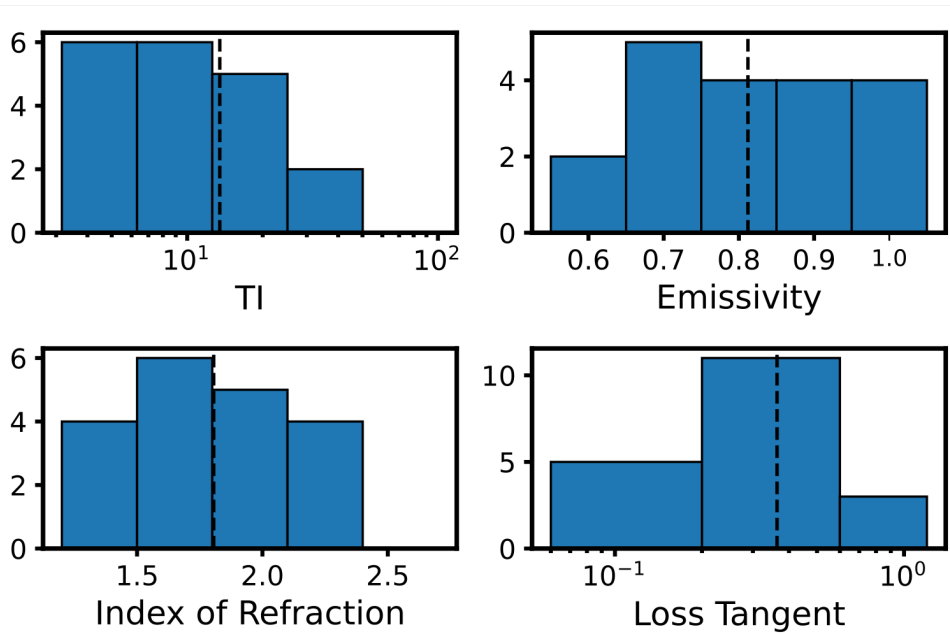

**Figure 3.** The distributions of all 19 best-fit parameters for Juno that are statistically acceptable and indistinguishable in terms of the goodness of fit to the data. The dashed vertical lines mark the averages. TI in the top left panel represents thermal inertia.

# 3. Discussion
## 3.1. Physical properties of Juno's regolith

In order to interpret the model fitting results, we turned to the laboratory measurements of t he thermal and dielectric parameters of the Juno surface analog materials. Juno is an Sq-type asteroid (DeMeo et al. 2009). S-type asteroids are generally considered to be associated with ordinary chondrites (OC) based on their spectra (e.g., Chapman and Salisbury 1973, Moroz et al. 1996, Eschrig et al. 2022) and confirmed by sample return (Nakamura et al., 2011). Noonan et al. (2019) compared Juno surface reflectance spectra (λ=0.7–2.55 um) to spectra H, L, and LL types of OCs, and determined that it is dynamically feasible for Juno (89% probability) to be the parent body for H-type OCs (Noonan et al. 2024).

However, the laboratory characterization of OCs for both dielectric and thermal physical properties at mm wavelengths and appropriate particle sizes for a regolith is very limited. Although the dielectric parameter measurements of a wide range of minerals exist for frequencies up to 190 GHz (e.g., Campbell & Ulrichs 1969, Hickson et al. 2020, Brouet et al. 2014), most of them correspond to lunar or carbonaceous materials. Brin et al. (2022) provides the measurements for OC materials, but only covers frequencies up to 1 GHz. On the other hand, the Microwave Instrument for Rosetta (MIRO) acquired observations of the M-type asteroid (21) Lutetia at a similar wavelength range to the Juno data from ALMA. Thermal and dielectric modeling of the data found the surface of Lutetia was consistent with the lunar regolith (Keihm et al. 2012, Gulkis et al. 2012). This is consistent with the result that the physical properties of regolith may dominate the thermal and dielectric properties over composition by Keihm et al. (2013). Also, Brouet et al. (2014) suggested that the real dielectric constant of particulate materials remains within 2 – 4 at frequencies up to 190 GHz. Therefore, here we use the limited laboratory measurements of OCs at lower frequencies (Brin et al. 2022) than our data to interpret the model results, and whenever unavailable, we turn to those of lunar regolith values (e.g., Carrier et al. 1991) to derive the physical properties about Juno's regolith.

### 3.1.1 Regolith density and porosity

Based on laboratory work, the density of multiphase mixture solely depends on the index of refraction, following the Lichtenecker formula for the mixing dielectric materials (e.g., Zakri et al. 1998),

$$\rho = 2 \frac{\ln n}{\ln \rho_0} \qquad (5)$$

where $\rho$ is in g cm$^{-3}$, $n$ is the index of refraction, and $\rho_0$ takes 1.938 g cm$^{-3}$ based on OC samples (Brin et al. 2022). Therefore, our modeled index of refraction could be directly converted to regolith density of 1.8 g cm$^{-3}$, which translates to a porosity of about 45% porosity assuming the OC grain density of 3.3 g cm$^{-3}$ (MacLennan and Emery 2022).

### 3.1.2 Thermal inertia and grain size

Our averaged modeled thermal inertia of a few to ~20 tiu is lower than the most recently reported value of $\Gamma = 40 - 85$ tiu based on photometric data from the Herschel Space Telescope at λ = 70 – 160 μm (Alí-Lagoa et al. 2020) using a similar shape model as what we use in this work. On the other hand, some thermal inertia values in the 19 statistically acceptable solutions (Table A1) are actually consistent with the previously derived range. Therefore, our results are considered consistent with previous measurements. The derived thermal inertias and densities of Juno's regolith indicate thermal skin depths between $\delta = 0.4$ mm and $\delta = 3.5$ mm.

Thermal inertia depends on thermal conductivity, $\kappa$, specific heat, $c$, and bulk density, $\rho$, (Section 2). Both $\kappa$ and $c$ are temperature dependent. While in the thermal models we assumed that their parameters are independent of temperature, interpretation of the best-fit thermal inertia does require consideration of the mean temperature. Following MacLennan and Emery (2022), the specific heat of S-, V-, and C-type asteroid can be expressed as,

$$c = 3.39T - 0.0033T^2 \qquad (6)$$

where $T$ is the temperature in K.

For thermal conductivity, we used the models from Gundlach and Blum (2012) and Sakatani et al. (2017). In both models, thermal conductivity is composed of two terms that represent heat conduction through grains and thermal radiation through pores, respectively,

$$\kappa = \kappa_1 + \kappa_2 T^3 \qquad (7)$$

The heat conduction through grains depends on the contact radius between grains, $r_c$. The model by Gundlach and Blum (2013) describes $\kappa_{1,GB}$ as,

$$\kappa_{1,GB} = 2.12 \times 10^{-2} \kappa_{grain} \frac{r_c}{r_g} \exp(5.26\psi) \qquad (8)$$

where $\kappa_{grain}$ is the thermal conductivity of a grain with zero porosity, $r_g$ is the grain radius, and $\psi = 1 - \phi$ is the volume filling factor for a porosity $\phi$. The constants are all empirically derived. The contact radius is,

$$r_c = \left[\frac{9\pi}{4} \frac{(1-\nu^2)}{E} \gamma(T) r_g^2\right]^{1/3} \qquad (9)$$

where $E$ is Young's modulus, and $\nu$ is Poisson's ratio, and $\gamma(T) = 6.67 \times 10^{-5} T$ in units of J m$^{-2}$ is the specific surface energy of each grain.

In Sakatani et al. (2017), the solid grain conductivity is modeled as,

$$\kappa_{1,S} = 0.4\, \kappa_{grain} \frac{r_c}{r_g} \frac{4\psi C}{\pi^2} \qquad (10)$$

The coordination number,

$$C = \frac{2.8112 \psi^{-1/3}}{f^2(1+f^2)} \qquad (11)$$

where $f = 0.07318 + 2.193\phi - 3.357\phi^2 + 3.914\phi^3$. The contact radius follows Eq. 9.

The radiative thermal conductivity coefficient, $\kappa_2$, for Gundlach and Blum (2013) and Sakatani et al. (2017) are expressed in the following equations, respectively,

$$\kappa_{2,GB} = 10.72\ \epsilon\sigma_0 \frac{\phi}{\psi} r_g \qquad (12)$$

$$\kappa_{2,S} = 13.6\ \sigma_0 \frac{\epsilon}{2-\epsilon} \left[\frac{\phi}{\psi}\right]^{1/3} r_g \qquad (13)$$

where $\epsilon$ is the bolometric emissivity, and $\sigma_0$ is Stefan-Boltzmann constant.

Fig. 4 shows those two models using S-type asteroid values (MacLennan and Emery 2022). The parameters we used are listed in Table 1, if not already mentioned above. These models suggest that a low thermal inertia of ~10 tiu as we derived from the ALMA data requires a high porosity of 70 – 90% and a small grain size of 10s µm. Such a high porosity has been reported for other objects. For example, Ceres's regolith could have a porosity of >70% (Palmer et al. 2021). Fine-grained regolith with 10s of µm grain size is also not uncommon on large asteroids or other planetary bodies (e.g., McKay et al. 1991, Delbo et al. 2015).

However, the index of refraction as derived from the ALMA data suggests a much lower porosity of ~45%, which is inconsistent with these models for any grain sizes. Fig. 5 shows the minimum thermal inertia allowed by the two models with respect to density, which directly relates to porosity. Most thermal inertia and density combinations are below the allowed range for both models that we calculated, unless the surface materials have a low thermal conductivity (Fig. 5a). The curves corresponding to different temperatures in Fig. 5b suggest that this result is not sensitive to the exact temperature that we modeled for Juno.

If the moderate porosity indicated by the index of refraction is correct, then a mechanism would be needed to explain the low observed thermal inertia. One possible mechanism might be the reduced particle contact area that can potentially lower the thermal conductivity. The Johnson et al. (1971) relationship for particle contact conduction (used by the thermal conductivity models we adopted in this work) includes a term for external forces. If the particles were to have some kind of repulsive forces, such as electrostatic charges, then that would reduce the effective surface energy, thus reducing the thermal conduction between particles. In the extreme case of the some repulsive force canceling the van der Waals attraction, the thermal conduction term goes to zero, and the model thermal inertia could potentially match the thermal inertia derived from the ALMA data while maintaining ~45% porosity for 10s of μm grain size (Fig. 6).

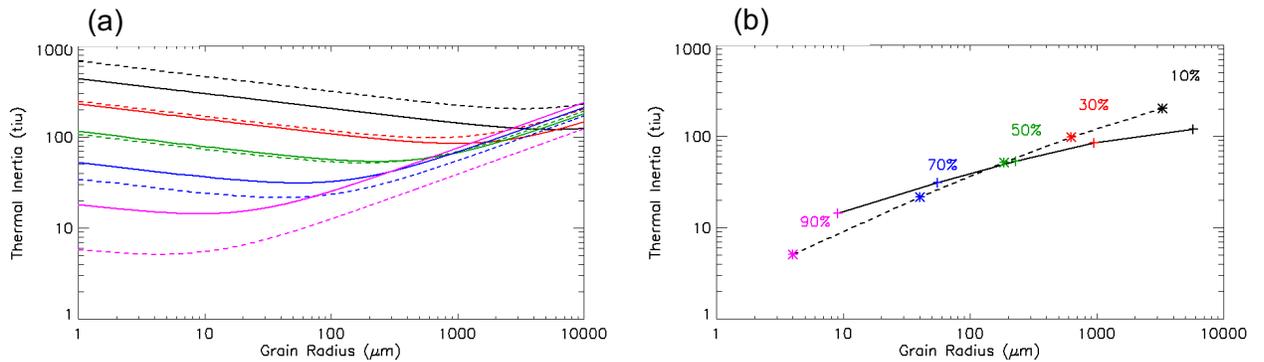

Figure 4. Modeled thermal inertia with respect to grain size for S-type asteroid materials. Panel (a) shows the dependence, and panel (b) shows the minimum possible thermal inertia for given porosities as marked in the plot. The curves of different colors from top to bottom represent different porosities: black 10%, red 30%, green 50%, blue 70%, and magenta 90%. The solid curves are based on Gundlach and Blum (2013), and the dashed curves are based on Sakatani et al. (2017).

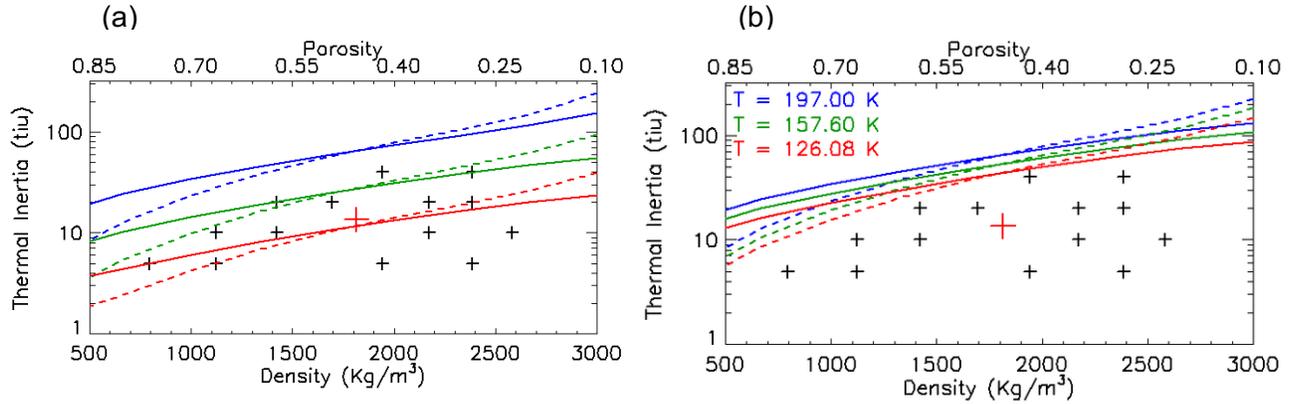

Figure 5. Minimum thermal inertia allowed by the Gundlach and Blum (2013) model (solid line) and the Sakanati et al. (2017) model (dashed line) with respect to density (lower axis), which represents porosity (top axis), using the S-type asteroid material parameters. (a) Blue: thermal conductivity k = 4.05 W / (m K); green: k = 0.405 W / (m K); red: k = 0.0405 W / (m K). The grain density is assumed to be $3.3\times10^3$ kg/m$^3$. (b) The colors represent three surface temperatures to test the sensitivity of the thermal inertia model with respect to temperature. A thermal conductivity of k = 4.05 W / (m W) is used in this panel, corresponding to the blue curves in panel (a). Below the curves is the region not allowed by models. The small black pluses in both panels mark the top 19 best-fit values (see text), and the red plus indicates the weighted average values. Most fitted thermal inertia values are lower than the model allowed, unless the material has a low thermal conductivity.

Table 1. Parameter values used in our thermal inertia model.

|  | L5 OC/S-type | Reference | Units |
|---|---|---|---|
| $\rho_g$ | 3331 | MacLennan and Emery (2022) | kg m$^{-3}$ |
| $\nu$ | 0.14 | Ibrahim (2012) used by MacLennan and Emery (2022) | unitless |
| E | 28.8e9 | Ibrahim (2012) used by MacLennan and Emery (2022) | GPa |
| $\kappa_g$ | 4.05 | MacLennan and Emery (2022) | W m$^{-1}$ K$^{-1}$ |

In addition to fine-grained regolith, recent work suggested that fractures or porous structure could also efficiently reduce thermal inertia of rocks (Cambioni et al. 2019, Lucchetti et al. 2021, Rozitis et al. 2020, Okada et al. 2020). However, those results were based on the observations of small asteroids of km or smaller in diameter, whose surfaces are dominated by exposed rocks. For asteroids as large as Juno, their surfaces are dominated by a layer of regolith with a limited abundance of exposed rocks (e.g., Noma and Hirata 2024, Schröder et al. 2021, Küppers et al.

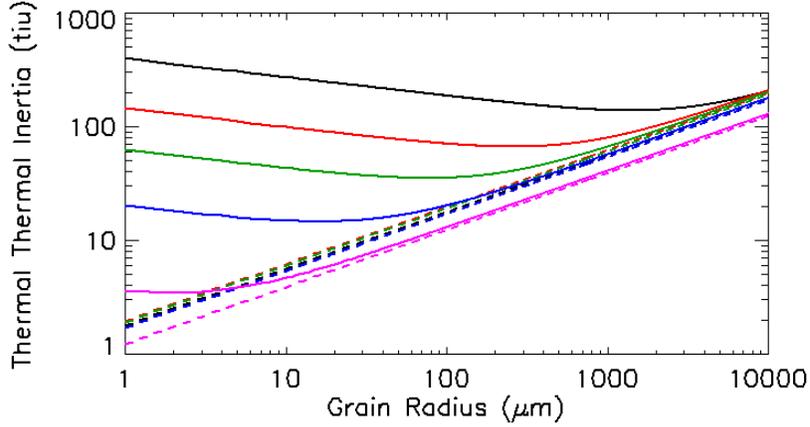

Figure 6. Modeled thermal inertia with respect to grain size radius. The different colors reflect changes in porosity, same as in Fig. 4. The solid lines represent the Sakatani et al. (2017) model using lunar-basaltic physical properties, while the dashed line represent the same model without the grain-contact conduction.

2012). Therefore, fractured or highly porous rocks might not be a dominant factor to contribute to the low thermal inertia for Juno.

### 3.1.3 Loss tangent

Given the assumption of $\tan \Delta \ll 1$ for Eq. 2, we need to ask how much the fitted loss tangent of up to 1 is still meaningful. In fact, even for a loss tangent of 1, the square root of $K_{er}$ and $n$ would differ by only ~10%. For a loss tangent of 0.31 and 0.1, the differences are ~2% and ~0.1%, respectively. Therefore, the best-fit loss tangent values that we derived are still reasonably accurate.

The dielectric loss tangent derived from the model fitting is indicated by the temperature at the depth probed by the ALMA observations at $\lambda = 1.3$ mm. Because the subsurface temperature profile depends on the thermal skin depth, which in turn depends on the assumed $\rho$ and $c$ in our model fitting (Eq. 1), we need to correct the fitted loss tangent for the actual estimated $\rho$ and $c$ for Juno, such that the results are self-consistent. This correction can be done by directly scaling the dielectric skin depth, or the dielectric absorption coefficient, by thermal skin depth, which inversely scales with $\rho c$ (Eq. 1),

$$\mu_{corr} = \mu \frac{\rho c}{\rho_{krc} c_{krc}} \quad (14)$$

here $\rho_{krc}$ and $c_{krc}$ are the density and specific heat capacity assumed in our thermophysical modeling. The corrected loss tangent for Juno's regolith is between 0.09 and 1.3 (Table A1).

The derived dielectric parameters of Juno indicate an electric absorption length $L = 0.1$ to 1.4 mm, which is shorter than or comparable to the wavelength of the observations and about a half of the thermal skin depth, $\delta$. Given the low thermal inertia, this is consistent with the gross correlation between Juno's brightness temperature at this wavelength with the subsolar point (ALMA Partnership et al. 2015). On the other hand, the loss tangent that we derive is much higher than the values commonly adopted for dry, silicate regolith, such as the Moon's (0.01; Gary &

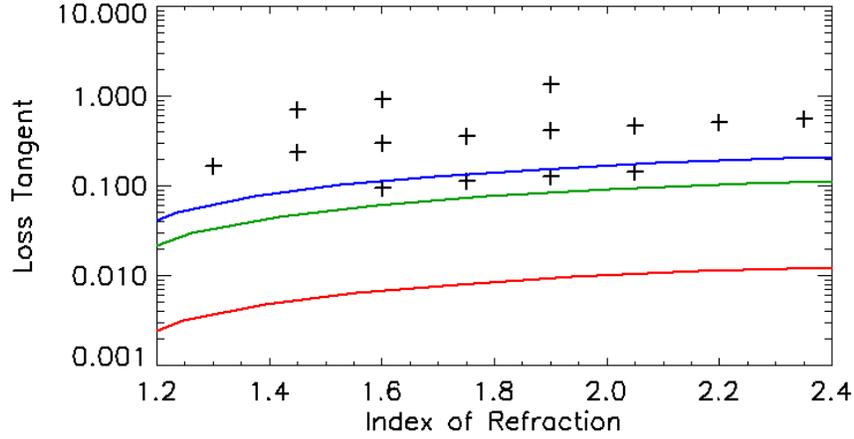

**Figure 7.** Derived loss tangent for Juno (symbols) compared to the values predicted by models (curves) with L5 OC material properties. The red curve is based on Brin et al. (2022), the green curve is from Brin et al. (2022) model using the average dielectric values from Garvin et al. (1988), and the blue curve is from Brin et al. (2022) model using Garvin et al. (1988), assuming the minimal range for the real component and the maximum range for the imaginary component. The fitted loss tangent is high compared to the model predictions.

Keihm 1978) and other S-type asteroids' (e.g., Gulkis et al. 2012), but close to the value of Ceres determined from ALMA data at a similar wavelength (Li et al. 2020). The high loss tangent value is also consistent with the relatively high brightness temperature of Juno near $\lambda = 1$ mm compared to other wavelengths (Redman et al. 1990; see Section 3.3).

We compared the derived loss tangent for Juno with the laboratory measurements (Fig. 7) based on Brin et al. (2022). Even using the minimum range for the real component and the maximum range for the imaginary component provided by Garvin et al. (1998), the loss tangent for Juno is still mostly higher than the model predictions. There might be some yet-to-be recognized mechanisms at work that increase the dielectric loss in the regolith of Juno.

### 3.1.4 Surface roughness

Following Capria et al. (2014), assuming an intrinsic gray body bolometric emissivity for a smooth surface $\varepsilon_0 = 0.95$, then our best-fit emissivity, $\varepsilon_{eq}$, can be related back to the standard IR beaming factor used in the Standard Thermal Model (STM; Lebofsky et al. 1986),

$$\eta = \frac{\epsilon_{eq}}{\epsilon_0} \qquad (15)$$

Furthermore, the roughness parameter, $\xi$, is calculated following Eq. (16),

$$\varepsilon_{eq} = (1 - \epsilon_0 \xi)\epsilon_0 = \eta \epsilon_0 \qquad (16)$$

The equivalent emissivity that we fit is therefore converted to an IR beaming parameter of 0.84, which is close to the typical values for main-belt asteroids (~0.79, with a FWHM of the distribution ~0.4; Masiero et al. 2011). The roughness parameter $\xi$ is derived to be ~0.2. If one populated the surface with hemispheric craters that are below the image resolution but larger than the thermal skin depth, roughly 24% of the surface would need to be covered for this equivalent roughness. This is reasonable, but lower than that of Lutetia, which would need ~50% of its surface covered by craters to match the data by comparing VIRTIS and MIRO observations (Keihm et al. 2012). The slightly lower roughness of Juno is also consistent with its slightly higher beaming parameter than the typical values of main-belt asteroids.

## 3.2. Rotational variations of regolith properties

ALMA Partnership et al. (2015) suggested possible longitudinal variations in the regolith properties as indicated by the shift of the hottest point on Juno from the afternoon side in early images (rotational phase $\varphi = 0.33 - 0.58$ as defined in ALMA Partnership et al. 2015) to close to the subsolar point in later images ($\varphi = 0.62 - 0.89$). To investigate such putative variations in the regolith properties, we calculated the ratio of the observations to the best-fit model assuming uniform regolith properties (Fig. 8) and simulated the thermal images as observed by ALMA (Fig. 9, also see supplementary video).

It appears that the early images with $\varphi = 0.33 - 0.58$ correspond to higher fluxes than the model prediction, and the later images with $\varphi = 0.62 - 0.89$ correspond to lower fluxes, although such variations are within about 1%. If this indicates a relatively high contribution from the thermal emission from shallower subsurface material for earlier images due, e.g., to high dielectric loss, then it should be associated with less thermal delay than that from deeper subsurface. Shorter thermal delays should result in the hottest spot being closer to local noon, contradicting the observations. Higher index of refraction could also decrease the sounding depth but could also result in lower millimeter emissivity because it increases the Fresnel reflection coefficient (Eq. 3). The net effect could depend on other parameters and need to be quantified. Alternatively, a lower thermal inertia could cause higher temperature at the same sounding depth, given that our modeled sounding depth is smaller than thermal skin depth. However, this scenario also indicates smaller thermal delays. On the other hand, a locally higher roughness, represented in our model by a lower equivalent emissivity $\varepsilon_{eq}$, could result in higher local surface temperature. The complex topography of Juno also plays a role in the location of the subsolar point with respect to the areas of peak temperature (Fig. 8). Therefore, we suggest that such a relative shift of the hottest spot on Juno could be either due to local topography or rotational variations in surface roughness or the combinational effect of them, but less likely due to rotational variations in the thermal inertia or loss tangent. It is difficult, though, to determine which parameter or combinations of parameters are responsible for the observed variations without detailed quantitative analysis with disk-resolved images (e.g., Cambioni et al. 2022). Our disk-integrated modeling does not allow us to definitively distinguish the effects of those different scenarios.

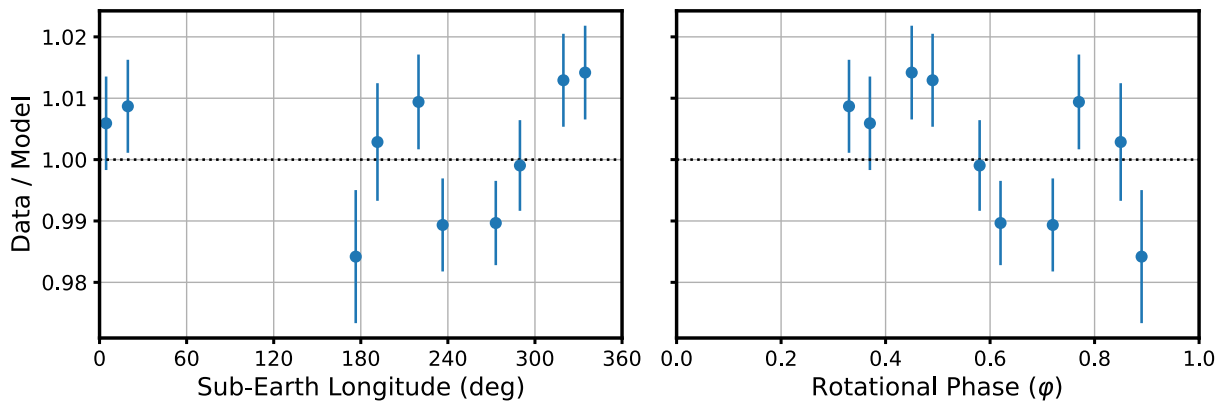

**Figure 8.** Ratios of the observed flux to the model for Juno, assuming uniform thermal and dielectric properties, with respect to longitude (left) and rotational phase, $\varphi$ (right). The rotational phase follows ALMA Partnership et al. (2015). The error bars reflect the uncertainty in the flux measurement, not including any systematic uncertainty in the thermal model.

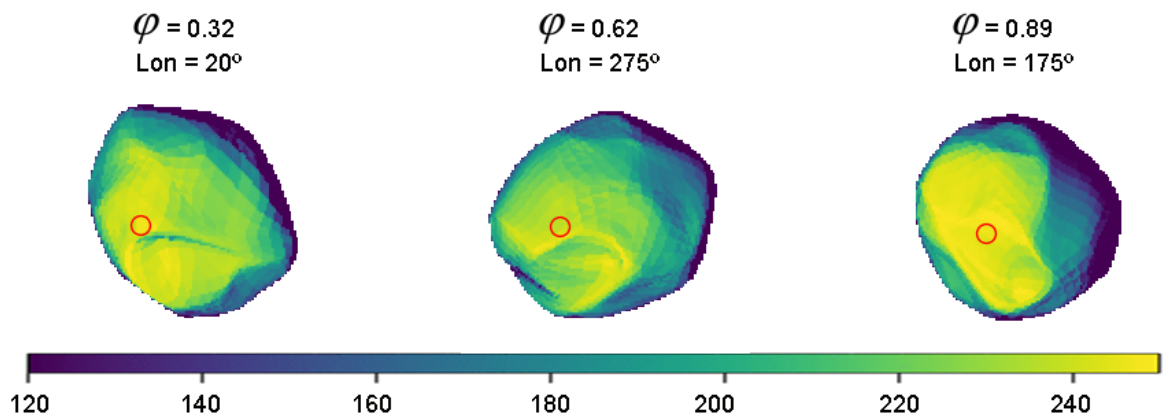

**Figure 9.** Simulated flux density distribution of Juno at λ = 1.3 mm at three rotational phases. The simulation uses thermal inertia Γ = 10 tiu, equivalent emissivity $\varepsilon_{eq}$ = 0.8, index of refraction $n$ = 1.75, and loss tangent tan Δ = 0.32. The color bar is in units of Jy arcsec$^{-2}$. The rotational phase follows the definition in ALMA Partnership et al. (2015). The red circle in each frame marks the location of sub-solar point. All frames have sky north in the up direction and east to the left. The relative shift between the subsolar point and the area with the highest temperature display a complex correspondence. See the associated animation in the Supplementary Material showing the flux density distribution during the rotation of Juno with a longitudinal sampling of 5º.

## 3.3. Effective emissivity and the 1-mm bump

Effective emissivity, $\varepsilon_{eff}$, is often used in thermal observations to characterize the ratio of the observed flux density to the modeled thermal flux density from the surface of an object based on STM (Lebofsky et al. 1986). Because mm wavelength observations probe subsurface thermal emission, which has lower temperature than the surface emission and is reduced by the Fresnel coefficient, the effective emissivities of asteroids at this wavelength are usually less than 0.8 (e.g., Redman et al. 1998).

If we use the average diameter of Juno (Viikinkoski et al. 2015), and assume A = 0.09 and an intrinsic gray body bolometric emissivity for a smooth surface $\varepsilon_0$ = 0.95, we find $\varepsilon_{eff}$ = 0.88±0.05 for Juno at λ = 1.3 mm. The uncertainty is dominated by the systematic flux calibration uncertainty of the ALMA data (ALMA Partnership et al. 2015). Our calculated value for Juno is much higher than other S-type asteroids at similar wavelengths, such as (7) Iris, (8) Hebe, and (18) Melpomene, and even more so than (4) Vesta (V-type) and M-type asteroids (16) Psyche and (216) Kleopatra (Redman et al. 1998). On the other hand, this emissivity is comparable to some measurements of Ceres at similar wavelengths (Redman et al. 1998). The high effective emissivity of Juno is also consistent with its high loss tangent as we derived, suggesting a shallower probing depth at this wavelength than at other wavelengths.

Comparing the spectral energy distribution (SED) of Juno between $\lambda$ = 0.2 and 1.2 mm (Redman et al. 1990) with other asteroids (e.g., Redman et al. 1998, Keihm et al. 2013), we note that Juno shows an increase in the brightness temperature around $\lambda$ = 1 mm. Because the electric penetration depth is proportional to wavelength, if the loss tangent remains constant, then as wavelength increases, the brightness temperature should decrease. Therefore, the increase of brightness temperature for Juno around $\lambda$ = 1 mm may be the result of increasing loss tangent at this wavelength. Comparing the spectral emissivity of Juno with other asteroids, we noticed that Vesta and other S-type asteroids (7 Iris, 8 Hebe, and 18 Melpomene) show decrease of effective emissivity from $\varepsilon_{eff}$ > 1 at $\lambda$ = 10 μm to $\varepsilon_{eff}$ = 0.6 – 0.8 at mm wavelengths, whereas M-type asteroids (16 Psyche and 216 Kleopatra) display a much stronger drop (Redman et al. 1998). On the other hand, Ceres and probably (10) Hygiea, both C-type with similar 3-μm spectral features (Rivkin et al. 2012, Takir & Emery 2012), show a much smaller drop of emissivity near $\lambda$ = 1 mm (Keihm et al. 2013) and are similar to Juno. This behavior is consistent with the high loss tangent that we derived for Juno here, as well as for Ceres (Li et al. 2020).

The spectral characteristics of asteroids at $\lambda$ ~ 1 mm are hard to explain. The similar behaviors within the C-type, S-type (except for Juno), and M-type asteroids seem to suggest that composition controls the spectral emissivity properties of asteroids, which could be ascribed to the different dielectric properties of various minerals. However, the exception of Juno in the S-types, and its similarity with the C-types, suggests that the physical structure of regolith could also make significant contributions in determining the dielectric properties. Laboratory measurements of the dielectric properties of a variety of samples with various physical conditions over frequencies into the 100s GHz range will be key to solving this "1 mm bump" puzzle.

## 4. Conclusion

In this work, we modeled the disk-integrated thermal lightcurve of Juno at $\lambda$ = 1.3 mm measured by ALMA covering 60% of its rotation. A thermophysical model taking into account the shape model of Juno was used to calculate the surface and subsurface temperature, and a radiative transfer model was used to account for subsurface thermal emission. We derived a set of thermal and dielectric parameters of Juno's regolith. The thermal inertia of Juno is constrained to be $\Gamma$ = 13±10 tiu, and the statistically acceptable values compatible with the data range between 5 and 40 tiu, which is consistent with previous measurements in the thermal infrared. The equivalent emissivity is 0.8±0.1, the index of refraction is 1.8±0.3, and the loss tangent is 0.4±0.3, although caveats exist about the reported uncertainties.

Based on the thermal and dielectric parameters derived from the ALMA data, we can infer some physical properties of Juno's regolith. Based on the laboratory measurements of the dielectric properties of meteoritic samples, the modeled index of refraction indicates a porosity of about 45%. However, using the laboratory measurements of the thermal properties of meteoritic samples, the models by Gundlach and Blum (2013) and by Sakatani et al. (2017) suggest that the Juno's surface should have a thermal inertia higher than the measured values, unless the porosity is >70%.

Therefore, there is probably some mechanisms acting in Juno's regolith that reduces the thermal contact between particles and therefore decrease the thermal conductivity. The loss tangent of Juno derived from the model and corrected for the actual density of the surface is likely in the order of 0.5, which is much higher than laboratory measurements for OCs (Brin et al. 2022), albeit at a much higher frequency than the latter. The derived beaming parameter and surface roughness of Juno are consistent a lower roughness than typical main-belt asteroid.

The effective emissivity of Juno at $\lambda = 1.3$ mm is about $\varepsilon_{eff} = 0.88\pm0.05$, consistent with its high loss tangent but much higher than other S-type asteroids or lunar regolith at similar wavelengths. Deviation of the observations from a uniform surface model suggests possible thermal and/or dielectric properties on the surface of Juno. On the other hand, local topography is the likely factor to cause the shift of the hottest spot relative to the sub-solar point as Juno rotates as noticed earlier (ALMA Partnership et al. 2015).

Comparing the SEDs of asteroids of various taxonomic groups found in the literature, C-type asteroids such as Ceres and possibly Hygiea show a "bump" in the brightness temperature at $\lambda \sim 1$ mm, whereas S-type asteroids do not show this behavior except for Juno. This may indicate that the spectral behavior of the dielectric loss tangent of asteroid regolith is controlled by mineralogic composition, but physical structure such as porosity could potentially make a significant contribution. Laboratory measurements of the dielectric properties of asteroid regolith analogs with a variety of mineralogic compositions under various physical conditions into the 100s GHz regime are key to improving our understanding of the ALMA observations of asteroids.

## Acknowledgments

This research is partially supported by NASA's Solar System Observations (SSO) R&A Program through grant NNX15AE02G to the Planetary Science Institute. J.-Y.L. and H.H.H. acknowledge partial support from the Solar System Exploration Research Virtual Institute 2016 (SSERVI16) Cooperative Agreement (grant NNH16ZDA001N), SSERVI-TREX to the Planetary Science Institute. This paper makes use of the following ALMA data: ADS/JAO.ALMA#2011.0.00013.SV. ALMA is a partnership of ESO (representing its member states), NSF (USA) and NINS (Japan), together with NRC (Canada), MOST, and ASIAA (Taiwan), and KASI (Republic of Korea), in cooperation with the Republic of Chile. The Joint ALMA Observatory is operated by ESO, AUI/NRAO, and NAOJ. The National Radio Astronomy Observatory is a facility of the National Science Foundation operated under cooperative agreement by Associated Universities, Inc. This research made use of a number of open-source Python packages (in arbitrary order): Astropy (Astropy Collaboration 2013), Matplotlib (Hunter 2007), SciPy (Virtanen et al. 2020), IPython (Pérez & Granger 2007), Jupyter notebooks (Kluyver et al. 2016). Any use of trade, firm, or product names is for descriptive purposes only and does not imply endorsement by the U.S. Government. We thank the anonymous reviewer for the comprehensive, detailed, and constructive review that has helped us improved this manuscript.

Facilities: ALMA

## ORCID IDs

Jian-Yang Li (李荐扬): https://orcid.org/0000-0003-3841-9977
Timothy N. Titus: https://orcid.org/0000-0003-0700-4875
Henry H. Hsieh: https://orcid.org/0000-0001-7225-9271
Arielle Moullet: https://orcid.org/0000-0002-9820-1032

# Supplementary Materials

**Table A1**. The top 19 best-fit parameters and the derived parameters following the approach described in Section 3.1. A few solutions result in an emissivity of 1, which lead to negative roughness parameter and are not included in this table.

| | Model Parameters | | | | | | Derived Parameters | | | | Skin Depths (mm) | |
|---|---|---|---|---|---|---|---|---|---|---|---|---|
| Reduced Chi-Squared | RMS (mJy) | Rel. RMS (%) | Thermal Inertial (tiu) | Emissivity | Index of Refraction | Loss Tangent | Density (kg/m3) | Corrected Loss Tangent | Beaming Parameter | Roughness Parameter | Dielectric | Diurnal Thermal |
| 2.02 | 1.88 | 0.96% | 5 | 0.7 | 1.900 | 0.316 | 1.88E+03 | 0.395 | 0.737 | 0.277 | 0.278 | 0.434 |
| 2.06 | 1.9 | 0.97% | 5 | 0.9 | 1.451 | 0.316 | 1.11E+03 | 0.229 | 0.947 | 0.055 | 0.622 | 0.749 |
| 2.23 | 1.9 | 0.97% | 5 | 0.6 | 2.200 | 0.316 | 2.30E+03 | 0.487 | 0.632 | 0.388 | 0.197 | 0.353 |
| 2.37 | 1.94 | 0.99% | 10 | 1 | 1.451 | 0.316 | 1.11E+03 | 0.229 | 1.053 | | 0.622 | 1.497 |
| 2.58 | 2 | 1.02% | 10 | 0.6 | 2.350 | 0.316 | 2.49E+03 | 0.529 | 0.632 | 0.388 | 0.171 | 0.652 |
| 2.71 | 1.99 | 1.01% | 5 | 0.8 | 1.900 | 1.000 | 1.62E+03 | 1.079 | 0.842 | 0.166 | 0.111 | 0.434 |
| 2.75 | 2.05 | 1.05% | 20 | 0.9 | 1.750 | 0.316 | 1.65E+03 | 0.345 | 0.947 | 0.055 | 0.345 | 1.990 |
| 2.76 | 2.29 | 1.17% | 10 | 0.8 | 1.601 | 0.100 | 1.42E+03 | 0.093 | 0.842 | 0.166 | 1.374 | 1.184 |
| 2.86 | 2.17 | 1.11% | 10 | 0.9 | 1.601 | 0.316 | 1.39E+03 | 0.289 | 0.947 | 0.055 | 0.448 | 1.184 |
| 2.91 | 2.17 | 1.10% | 10 | 0.7 | 2.050 | 0.316 | 2.10E+03 | 0.443 | 0.737 | 0.277 | 0.231 | 0.776 |
| 2.92 | 2.18 | 1.11% | 40 | 0.8 | 1.900 | 0.100 | 1.93E+03 | 0.128 | 0.842 | 0.166 | 0.848 | 3.471 |
| 2.97 | 2.17 | 1.10% | 20 | 0.9 | 1.601 | 0.100 | 1.42E+03 | 0.093 | 0.947 | 0.055 | 1.374 | 2.369 |
| 3.02 | 2.32 | 1.18% | 20 | 0.8 | 1.750 | 0.100 | 1.69E+03 | 0.111 | 0.842 | 0.166 | 1.056 | 1.990 |
| 3.03 | 2.12 | 1.08% | 10 | 1 | 1.601 | 1.000 | 1.23E+03 | 0.787 | 1.053 | | 0.174 | 1.184 |
| 3.13 | 2.15 | 1.09% | 20 | 0.7 | 2.200 | 0.316 | 2.30E+03 | 0.487 | 0.737 | 0.277 | 0.197 | 1.413 |
| 3.17 | 2.25 | 1.14% | 40 | 0.7 | 2.200 | 0.316 | 2.30E+03 | 0.487 | 0.737 | 0.277 | 0.197 | 2.826 |
| 3.17 | 2.23 | 1.13% | 5 | 1 | 1.451 | 1.000 | 9.95E+02 | 0.625 | 1.053 | | 0.236 | 0.749 |
| 3.18 | 2.21 | 1.12% | 20 | 0.7 | 2.050 | 0.100 | 2.16E+03 | 0.143 | 0.737 | 0.277 | 0.703 | 1.552 |
| 3.19 | 2.2 | 1.12% | 5 | 1 | 1.301 | 0.316 | 7.85E+02 | 0.162 | 1.053 | | 0.976 | 1.059 |